\documentclass[10pt, conference, a4paper]{IEEEtran}
\IEEEoverridecommandlockouts
\usepackage{cite}
\usepackage{amsmath,amssymb,amsfonts}
\usepackage{hyperref}
\usepackage{algorithmic}
\usepackage{graphicx}
\usepackage{textcomp}
\usepackage{xcolor}
\usepackage{multirow}
\usepackage{adjustbox}
\usepackage{subfigure}
\usepackage{multicol}  % 用于创建多栏布局
\usepackage[normalem]{ulem}
\useunder{\uline}{\ul}{}

\def\BibTeX{{\rm B\kern-.05em{\sc i\kern-.025em b}\kern-.08em
    T\kern-.1667em\lower.7ex\hbox{E}\kern-.125emX}}

\begin{document}

\title{DRAMA: A Dynamic Packet Routing Algorithm using Multi-Agent Reinforcement Learning with Emergent Communication
}

\author{\IEEEauthorblockN{Anonymous Authors}}

\author{
\IEEEauthorblockN{Wang Zhang\IEEEauthorrefmark{1},  Chenguang Liu\IEEEauthorrefmark{1}\IEEEauthorrefmark{3}, Yue Pi\IEEEauthorrefmark{1}, Yong Zhang\IEEEauthorrefmark{2}, Hairong Huang\IEEEauthorrefmark{2}, Baoquan Rao\IEEEauthorrefmark{2}, \\ \footnotemark{\textsuperscript{1}}Yulong Ding\IEEEauthorrefmark{1}, \footnotemark{\textsuperscript{1}}Shuanghua Yang\IEEEauthorrefmark{1}\IEEEauthorrefmark{4}, \footnotemark{\textsuperscript{1}}Jie Jiang\IEEEauthorrefmark{1}\IEEEauthorrefmark{5}} 
  
\IEEEauthorblockA{\IEEEauthorrefmark{1}Department of Computer Science and Engineering, Southern University of Science and Technology, Shenzhen, China}
\IEEEauthorblockA{\IEEEauthorrefmark{2}Huawei Technologies Co., Ltd.} 
\IEEEauthorblockA{\IEEEauthorrefmark{3}School of Engineering, University of Warwick,
 Coventry, UK, CV4 7AL}
\IEEEauthorblockA{\IEEEauthorrefmark{4}Department of Computer Science, University of Reading, UK}
\IEEEauthorblockA{\IEEEauthorrefmark{5} China University of Petroleum (Beijing), Beijing, China}
\IEEEauthorblockA{Contact: 12232429@mail.sustech.edu.cn}
}
\maketitle
\footnotetext[1]{Corresponding author.}

\pagestyle{empty} %取消pagenumber
\thispagestyle{empty}

\begin{abstract}
The continuous expansion of network data presents a pressing challenge for conventional routing algorithms. As the demand escalates, these algorithms are struggling to cope. In this context, reinforcement learning (RL) and multi-agent reinforcement learning (MARL) algorithms emerge as promising solutions. However, the urgency and importance of the problem are clear, as existing RL/MARL-based routing approaches lack effective communication in run time among routers, making it challenging for individual routers to adapt to complex and dynamic changing networks.
More importantly, they lack the ability to deal with dynamically changing network topology, especially the addition of the router, due to the non-scalability of their neural networks. This paper proposes a novel dynamic routing algorithm, DRAMA, incorporating emergent communication in multi-agent reinforcement learning. Through emergent communication, routers could learn how to communicate effectively to maximize the optimization objectives. Meanwhile, a new Q-network and graph-based emergent communication are introduced to dynamically adapt to the changing network topology without retraining while ensuring robust performance. Experimental results showcase DRAMA's superior performance over the traditional routing algorithm and other RL/MARL-based algorithms, achieving a higher delivery rate and lower latency in diverse network scenarios, including dynamic network load and topology. Moreover, an ablation experiment validates the prospect of emergent communication in facilitating packet routing.

\end{abstract}

\begin{IEEEkeywords}
Dynamic Topology, Emergent Communication, Multi-Agent Reinforcement Learning, Packet Routing
\end{IEEEkeywords}

\section{Introduction}
Packet routing plays a vital role in computer networking, and it serves as a fundamental mechanism for efficient and reliable data transmission across sophisticated networking topologies. With the increasing amount of mobile users and services, wireless-connected distributed networks become one of the promising technologies to accommodate the vast requirements for robustness and adaptability to dynamic environments. This could be beneficial for various intelligent applications, such as vehicular ad-hoc networks (VANET) for road safety and traffic efficiency \cite{tang2019future}, unmanned aerial vehicle (UAV) swarms for disaster relief \cite{chen2020review}, and space-terrestrial integrated networks for global serving \cite{lai2023achieving}. Therefore, ensuring the quality of service (QoS) for packet routing in these networks, including latency and throughput, is of utmost importance. 

Several research works have been conducted to develop efficient routing algorithms. For example, traditional algorithms, such as OSPF \cite{moy1998ospf} and DSDV \cite{he2002destination}, use distance and manually specified metrics to offer a simplistic but efficient implementation. However, they may struggle to deal with scenarios with high levels of network congestion since the network conditions, such as traffic load and potential queuing delays, cannot be captured by them. On the other hand, reinforcement learning (RL), a sequential decision-making approach, has demonstrated great potential for packet routing \cite{stampa2017deep,reis2021r2l,jalil2020dqr,boyan1993packet,chen2020rl,almasan2022deep}. Some RL-based approaches assume a centralized controller \cite{stampa2017deep,jalil2020dqr,almasan2022deep}, where all agents' observations are concatenated into a single long input vector, and decisions are made simultaneously for each agent. They may face scalability challenges and optimization difficulties due to the vast action and state spaces. In contrast, the other RL-based approaches are distributed \cite{boyan1993packet,chen2020rl,reis2021r2l}, where each router is seen as an agent and treats other routers as part of the environment. However, the policy of each agent changes during the training phase. Thus, the individual agent often faces non-stationary issues \cite{oroojlooy2023review} with low cooperative ability using these methods.

Recently, multi-agent reinforcement learning (MARL) has emerged as an effective approach to addressing packet routing in distributed networks\cite{mukhutdinov2019multi,song2023trustworthy,pinyoanuntapong2019delay,you2020toward,chen2021multiagent,he2020deephop,li2020routing,qiu2022data,liu2021drl}. MARL-based algorithms allow routers to share their observations, actions, or rewards to reach a consensus on a globally optimized task, which could potentially alleviate network congestion and enhance packet delivery efficiency. However, most of these works \cite{mukhutdinov2019multi,song2023trustworthy,pinyoanuntapong2019delay,you2020toward,chen2021multiagent,he2020deephop} do not apply communication between routing nodes in deployment to reduce bandwidth utilization. Consequently, adapting to the dynamic changes in networks can be very challenging. In practice, distributed networks could demonstrate uncertainty and dynamic characteristics to accommodate extensive QoS requirements in networking \cite{tang2019future,chen2020review}, necessitating adaptability to network topology changes. Although some works in \cite{li2020routing,qiu2022data,liu2021drl} have been proposed to address link or router failures in MARL-based systems, they do not support the addition of new routers. Therefore, developing an adaptive MARL-based routing algorithm for dynamic link removal and node expansion in complex networks is critical.

On the other hand, with the continual development of multi-agent systems, emergent communication is proposed as a key approach to allowing collaborative agents to `learn to communicate' amongst themselves\cite{sukhbaatar2016learning}. Emergent communication has been applied to various applications, such as traffic control management\cite{gupta2020networked}, goal-oriented navigation\cite{das2019tarmac}, and autonomous driving\cite{yuan2023dacom}. By incorporating the transmission and aggregation of messages into a policy network in the multi-agent system, agents can learn what to communicate and with whom to communicate using end-to-end (E2E) training with task rewards. Compared with conventional communications with manually predefined communication protocols, the messages transmitted in emergent communication are vectors analogous to the hidden layers in neural networks, optimized by back-propagation. Although explaining this shared information is very challenging due to black-box optimization, it is highly informative, with task-specific features related to routing and networks.
Moreover, intelligent agents connected via emergent communication can self-organize their communication patterns based on their requirements. These capacities could facilitate negotiation and collaboration between agents, addressing real-time adaptation and scalability issues. This promisingly allows the MARL-based routing algorithm to collectively adapt to topology changes in complex networks. However, the application of emergent communication in packet routing has yet to be investigated. 

Motivated by these, we propose DRAMA, a dynamic packet routing algorithm using multi-agent reinforcement learning with emergent communication, to address the packet routing problem in dynamic networks. Our main contributions are summarized as follows:
\begin{itemize}
    \item  We propose an emergent communication-based MARL routing algorithm to address dynamic packet routing in MARL-based systems. The messages shared among routers are extracted from network states by neural networks before transmission and aggregated by an attention-based algorithm after reception. Different scenarios, including varying network loads and link bandwidths, are considered to validate its effectiveness and robustness. To the best of the authors' knowledge, this is the first work that uses emergent communication for dynamic packet routing. 

    \item Unlike aforementioned MARL-based works in \cite{li2020routing,qiu2022data,liu2021drl} that have neglected the route expansion in MARL-based routing system, we propose a combination of a novel Q-network and a graph-based emergent communication for dynamic node expansion and removal. Link/router failures and additional routers are considered for various QoS requirements, including packet delivery rate and latency.

    \item Numerical results show that DRAMA outperforms six baseline algorithms in synthetic and real-world networks, including the traditional, RL-based, and MARL-based algorithms. The proposed DRAMA is validated to effectively support dynamic router extension and removal without requiring retraining. Also, the network load experiments show that the proposed DRAMA could enable routers to make optimal collaborative decisions across various network statuses, 
    achieving the lowest latency. %%这句话需要吗
    Furthermore, an ablation study is conducted to demonstrate the effectiveness of emergent communication.
\end{itemize}

\section{Related Work}
\textbf{RL-Based Packet Routing Algorithm.} Q-routing \cite{boyan1993packet} was one of the earliest proposals for an RL-based packet routing algorithm. It treated each router as an independent agent, employing tabular Q-values to estimate packet transmission times. In \cite{stampa2017deep}, they proposed using deep reinforcement learning (DRL) to learn an effective mapping from observed demand matrices to routing strategies. DQR in \cite{jalil2020dqr} proposed an online SDN routing method. It utilized Dueling DQN and prioritized experience replay to learn static network topologies and optimized multiple E2E QoS metrics, including latency, bandwidth, lost rate, and cost. The work in \cite{almasan2022deep} integrated Graph Neural Networks (GNN) into RL policy networks of the central controller to prevent congestion in dynamic network topologies. However, most of these algorithms faced deployment and optimization challenges due to large state-action spaces, requiring a centralized controller. To address this, RL-Routing in \cite{chen2020rl} introduced a packet-level RL-based routing algorithm that deployed individual agents within each switch. It incorporated more comprehensive information, including switch throughput rate and link trust levels, to represent observations and enhance the agent's capacity for congestion management. The work in \cite{reis2021r2l} introduced an evolution strategy algorithm to optimize the policy function, which could dynamically adjust the transmission rate in packet routing. However, they encountered non-stationary problems and had difficulty learning stable policies.

\textbf{MARL-Based Packet Routing Algorithm.} To address the challenges mentioned above, the work in \cite{mukhutdinov2019multi} proposed the use of MARL to tackle the packet routing problem, employing a combination of Q-routing and Deep Q-network (DQN). Each router is treated as an agent with centralized training and decentralized execution(CTDE). Additionally, preliminary supervised learning was utilized for pre-training. Moreover, to reduce E2E latency, the work in \cite{pinyoanuntapong2019delay} employed MARL with a modular and composable learning approach, achieving lower latency through the composition of double learning, expected policy evaluation, and on-policy learning. In \cite{you2020toward}, a fully distributed MARL-based packet routing method, namely DQRC, was proposed, using a recurrent neural network for decisions in a fully distributed environment. Manually specified communication content, i.e., queue information, was transmitted between routers in DQRC, which is demonstrated to be more effective than no communication. Furthermore, to address time-varying traffic demand, the work in \cite{chen2021multiagent} incorporated meta-learning into MARL, optimizing with independent proximal policy optimization (PPO) \cite{schulman2017proximal}. To alleviate edge network congestion, DeepHop \cite{he2020deephop} utilized attention mechanisms to assist agents in comprehending the significance of each element within the network state. The work in \cite{song2023trustworthy} introduced a distributed trustworthy routing scheme based on the trust value proposed for low earth orbit (LEO) satellite networks, leveraging dueling double deep Q network to confirm load-balance well. However, these algorithms often lacked efficient communication between highly independent routers during the execution phase, failing to address dynamic topology issues.

DMARL in\cite{li2020routing} presented an efficient routing protocol for Underwater Optical Wireless Sensor Networks, addressing limited energy resources and highly dynamic topology by exchanging local/global reward and ACK packet. To adapt to topologically complex and dynamically changing networks in drone clusters, the work in \cite{qiu2022data} proposed a routing algorithm Based on multi-agent deep deterministic policy gradient (MADDPG \cite{lowe2017multi}) with LSTM as actor-network, extracting temporal continuity about all observations. Moreover, DRL-OR in \cite{liu2021drl} proposed a hop-by-hop method for scalability and a general utility function for multiple QoS requirements. Nevertheless, these dynamic routing approaches only communicated superficial messages and were limited in supporting router addition.

\textbf{Emergent Communication.}
To coordinate the behavior of each agent, CommNet in \cite{sukhbaatar2016learning} proposed a simple controller for MARL to learn to communicate among a dynamically varying group of agents through a fully connected communication channel. TarMac \cite{das2019tarmac} was the first to introduce attention mechanisms for weighted aggregation of received messages through other agents' keys, applying emergent communication to complex tasks such as goal-oriented navigation with continuously moving targets. To address the challenge of rapidly changing neighbors, DGN \cite{jiang2019graph} views the communication network as a graph instead of depending on broadcasting, utilizing graph convolution to aggregate messages and further improving cooperation through temporal regularization. In \cite{gupta2020networked}, the work introduced an algorithm for scenarios where agents communicate through fixed network connections in a discrete channel, applied explicitly in managing traffic controllers. Additionally, DACOM \cite{yuan2023dacom} investigated communication delay issues in emergent communication and applied it in an autonomous driving environment. Furthermore, the works in \cite{abudu2020deep} and \cite{chafii2023emergent} revealed the potential of applying emergent communication in IoT and future wireless networks. Nevertheless, these works have yet to explore the utilization of emergent communication in packet routing problems.

\section{System Model}
\subsection{Problem Formulation}
We formulate the packet routing optimization problem as a Decentralized Partially Observable Markov Decision Process (Dec-POMDP) constrained at the packet level, where each timestep $t$, each agent $i$ obtains a local partial observation $o^t_i$ from environment state $s^t$, takes an action $a^t_i$, and gets a reward $r_i^t$. With emergent communication, the agents should generate the messages $m^t_i$ according to $o^t_i$, then send them to others, and combine the received messages with observation to decide the action $a^t_i$. The goal for each agent is to determine an optimal policy $\pi_i^{*}: M_{-i} \times O_i \rightarrow M_i \times A_i$ such that its long-term discounted reward expectation (the value function) is maximized, where $O_i, A_i, M_i$ are observation, action, and message space of agent $i$ respectively, and $-i$ represents the indices of all the agents except agent $i$. The value function $Q^i$ of agent $i$ is determined by joint policy $\pi$ of all $N$ agents as $Q^i_{\pi^i,\pi^{-i}}(a,s):=\mathbb{E}[\sum_{t=0}^{\infty}\gamma^t \sum_{i=1}^{N}r_i^t(s^t, a_i^t)|_{a^t_i \sim \pi^i, a_0=a, s_0=s}]$, where $\gamma \in [0,1]$ is discount factor.

In networks, each router is considered as an intelligent agent. At each timestep, the collaborative routers are tasked with deciding the forwarding target of a data packet from the front of their queue so that all packets in networks can reach the destination successfully with averaged minimal latency. Moreover, we consider the network graph as an undirected graph $G=(V, E)$, where $V$ is the set of $N$ routers, and $E$ is the set of links between these routers. Hence, packet forwarding and emergent communication are constrained to occur only between neighbors in the graph.

\subsection{Action Space}
As mentioned above, each router is expected to choose one of its adjacent routers to transmit the specific packet. Thus, the action space $A_i$ of router $i$ will be the discrete set of neighboring agents denoted as $A_i=\{j|edge(i,j)\in E\}$. Notably, the dynamic nature of the number of adjacent router sets, affected by potential router/link failures or the addition of new routers, could result in corresponding changes in the action space. Such dynamism poses a significant challenge to the adaptability of Q-network in conventional RL/MARL-based algorithms.

\subsection{Observation Space}
The observation space $o_i$ of the router $i$ is determined by four parts as $o_i = \{o_{id}, o_h, o_q, o_d \}$. Specifically, $o_{id}$ is the unique identifier for breaking the symmetry and distinguishing each router individually. Additionally, $o_h$ serves as the action sequences of the router $i$ over the past steps, providing insights into the current network status. $o_q$ represents the information related to the router's queue buffer, including the queue length, maximum queue length, and the destination sequence of the packets in the queue. Lastly, $o_d$ is the degree of the router for subgraph representation.

\subsection{Reward}
The design of the reward function is crucial for the success of the RL/MARL algorithm, typically corresponding to the QoS. In this work, we are concerned with two key metrics: packet delivery rate and latency. Consequently, the reward $r_i$ of router $i$ forwarding a packet is given by:
\begin{equation}
 r_i = 
\begin{cases} 
-r_{\text{lost}} & \text{if the packet is lost}, \\
-r_{q} - r_{l} & \text{otherwise}.
\end{cases}
\label{eq:reward}
\end{equation}

Here, $r_l = \tau_l*t_l$ and $r_q=\tau_q*t_q$ are the penalties for the time cost of transmission through the neighboring link $t_l$ and queuing $t_q$ at the next-hop router, respectively, where $\tau_l$ and $\tau_q$ represent the penalty weight. Furthermore, the router is expected to incur a packet loss penalty $r_{lost}$ when the packet is transmitted to a router with a full queue. 

\begin{figure*}[t]
\centerline{\includegraphics[width=1\linewidth]{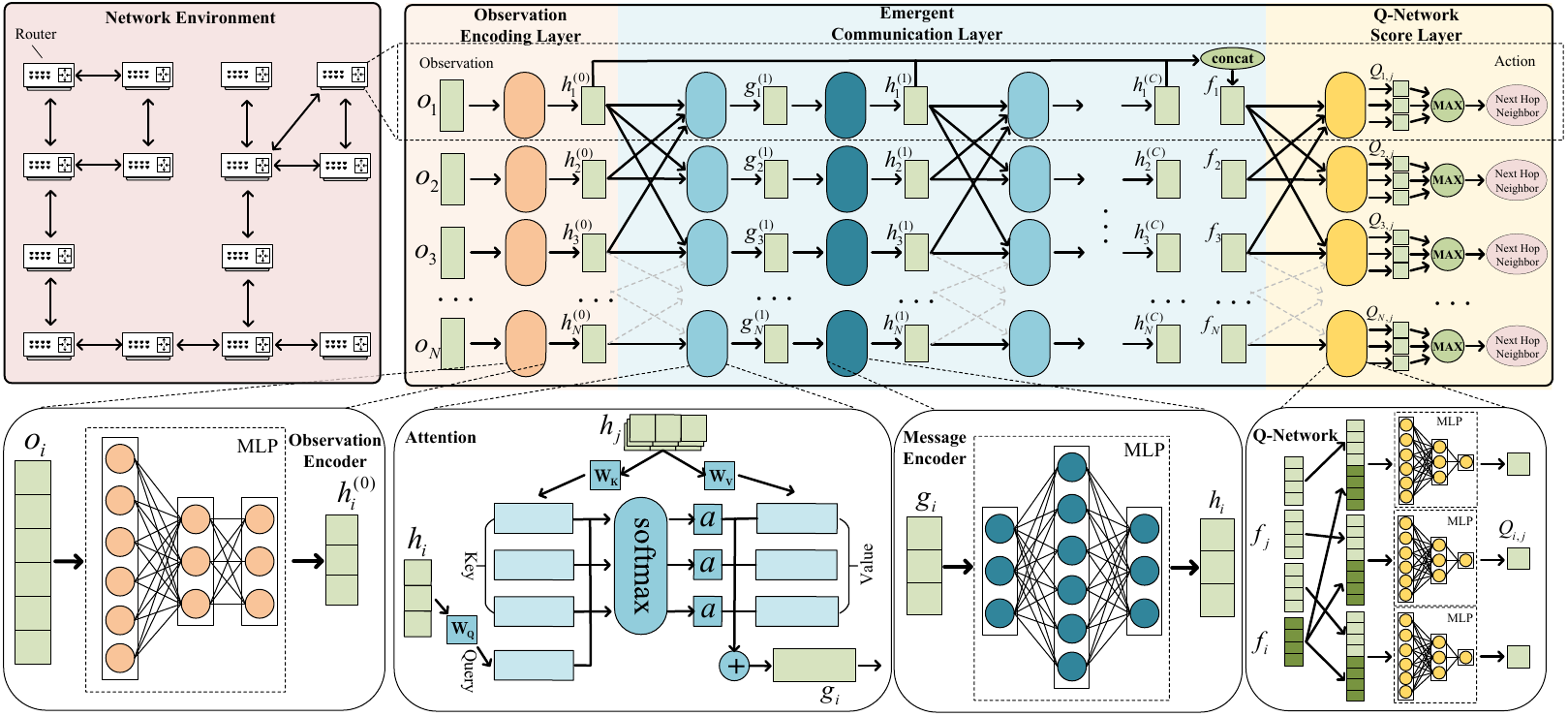}}
\caption{Overview of DRAMA}
\label{fig:framework}
\end{figure*}

\section{The proposed DRAMA}
\subsection{The Architecture}
To address the routing problem modeled as a Dec-POMDP, DRAMA is proposed for making packet forwarding decisions in routers. DRAMA comprises three key modules: observation encoding layer (OEL), emergent communication layer (ECL), and Q-network score layer (QSL), all composed of neural networks, as illustrated in figure \ref{fig:framework}. The observation encoding layer extracts crucial features from the observations. The emergent communication layer is responsible for repeatedly generating and sharing messages between routers. Notably, the messages exchanged in each communication round are not pre-coded, diverging from the conventional communication paradigm. As emergent communication, they are learned during training based on the optimization objective and encoded through neural networks in real time. The Q-network score layer evaluates the benefits of forwarding to neighbors. Our model distinguishes itself from other MARL-based packet routing algorithms due to the emergent communication layer and Q-network score layer, with the former layer fostering router collaboration, while the latter layer enhances the ability to adapt to dynamic topology challenges. Weights of DRAMA are shared amongst all routers to scale effectively.

\textbf{Observation Encoding Layer}. In the observation encoding layer, router $i$ takes the current observation $o_i$ as input, encoding the observation represented as the hidden state $h^{(0)}_i$, which becomes the input for the emergent communication layer. The observation encoding layer leverages a neural network as an observation encoder, such as the fully connected networks. Thus, $h^{(0)}_i = F_1(o_i)$.

\textbf{Emergent Communication Layer}. Inspired by DGN \cite{jiang2019graph}, our emergent communication layer utilizes graph-based emergent communication, which assists in extracting latent features from multiple hops to capture the structure and state of the local subnetwork. Specifically, agents engage in $C$ rounds of communication with their neighbors. In the $c$-th round, router $i$ sends the encoded message $h^{(c-1)}_i$ from the previous round to all neighboring routers. Note that $h^{(0)}_i$ is sourced from the observation encoding layer. Let $n_i$ be the set of indices for the neighbors of router $i$. Correspondingly, router $i$ receives messages $h^{(c-1)}_j$ from its neighbors, where $j \in n_i$. DRAMA leverages an attention mechanism for message aggregation, invariant to the number and order of neighbors, accommodating changing topology. Specifically, for router $i$, it uses its own hidden vector $h^{(c-1)}_i$ as a query and the messages $h^{(c-1)}_j$ from neighbors as keys and values. The attention coefficient is calculated as follows:
\begin{equation}
a_{ij}^{(c)} = \frac{\exp\left(\tau \cdot \mathbf{W}_Q h_i^{(c-1)} \cdot (\mathbf{W}_K h_j^{(c-1)})^\top\right)}{\sum_{k \in n_i} \exp\left(\tau \cdot \mathbf{W}_Q h_i^{(c-1)} \cdot (\mathbf{W}_K h_k^{(c-1)})^\top\right)}
\label{eq:agg1}
\end{equation}

Here, $\tau$ is a scaling factor, and $\mathbf{W}_Q$ and $\mathbf{W}_K$ are linear projection coefficients for query and key in the attention operation. After obtaining attention coefficients, agent $i$ combines all received messages as $g_i^{(c)}$ by weighted averaging according to the attention coefficient $a_{ij}^{(c)}$: 
\begin{equation}
g_i^{(c)} = \sum_{j \in n_i} a_{ij}^{(c)} \mathbf{W}_V h_j^{(c-1)}
\label{eq:agg2}
\end{equation}

Here, $\mathbf{W}_V$ is the linear projection coefficient for the value in the attention operation. Lastly, a non-linear neural network could be applied as the message encoder to generate the new message $h_i^{(c)}$ for the $(c+1)$-th round, as $h_i^{(c)} = F_2\left(g_i^{(c)}\right)$.

After $C$ rounds, the encoded messages from each communication are concatenated as the feature extracted by router $i$, denoted as $f_i$:
\begin{equation} f_i = \text{concat}[h^{(0)}_i, \ldots, h^{(C)}_i] 
\label{eq:message_enc}
\end{equation}

where $\text{concat}(\cdot)$ represents the concatenation function.
After $C$ communication rounds, router $i$ could acquire information from neighboring routers within the C-hop, engaging in communication and negotiation with more routers. 

Through graph-based emergent communication, the feature $f_i$ could learn both the topological structure of the neighborhood and the distribution of network-related features extracted by adjacent routers. It is worth noting that when a new router $w$ is introduced, communication with its neighbors in the original network is expected to assist in unveiling both its role and position within the local subgraph. This process aligns the feature space of $f_w$ with the existing routers.

\begin{figure}[t]
\centerline{\includegraphics[width=.8\linewidth]{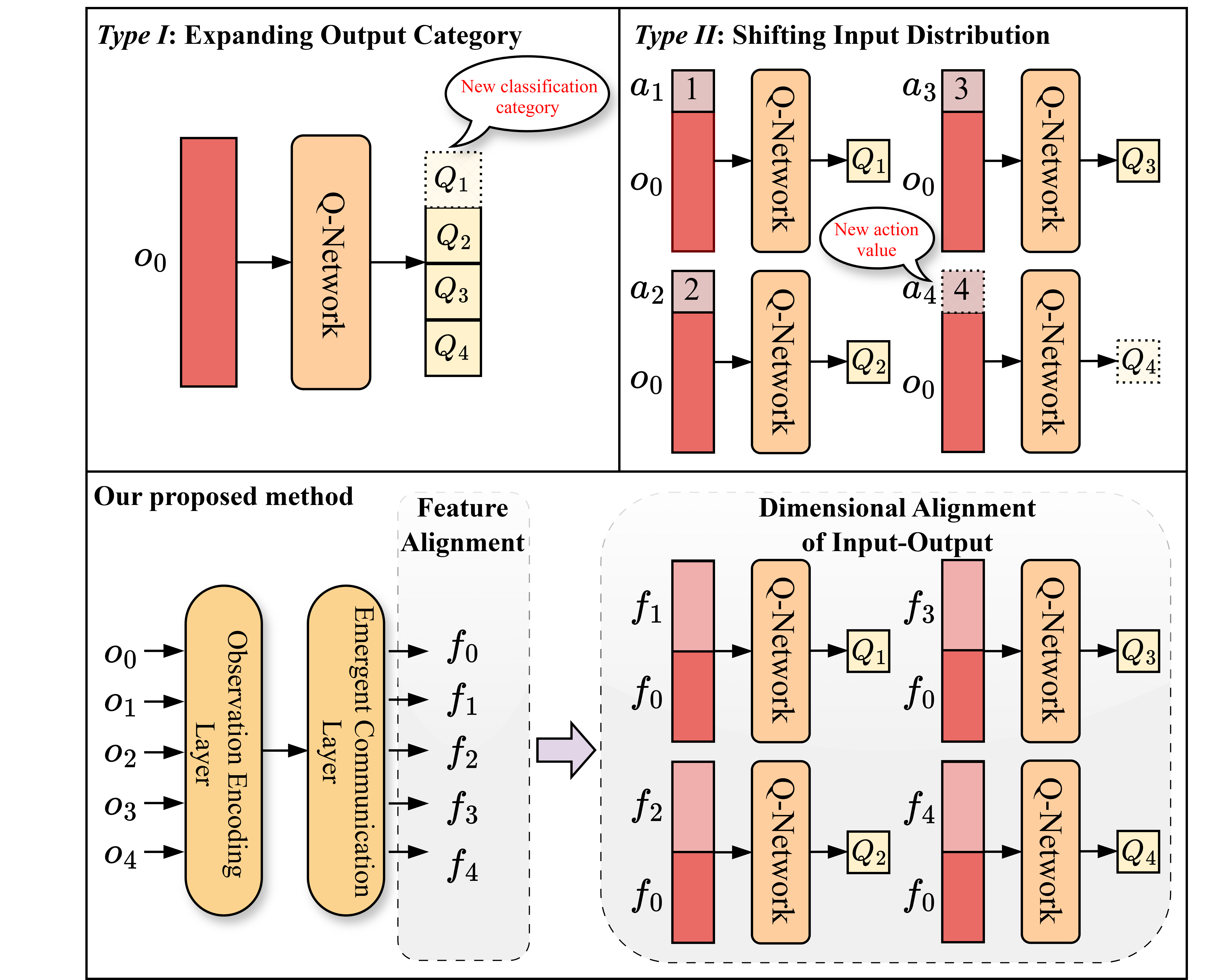}}
\caption{Two types of traditional Q-network and our proposed method. Router 0 calculates the Q-value for each neighbor. It has the original neighbors \{1,2,3\}, and the new neighbor 4. $o_i, a_i, f_i$ are observation, action, and feature of agent $i$, respectively, and we assume enumerative $a_i=i$.}
\label{fig:qnetwork}
\end{figure}

\textbf{Q-network Score Layer}. The Q-network score layer proposed in this paper differs significantly from traditional Q-networks to adapt to dynamic topology. In typical RL/MARL with discrete action spaces, the main task of a Q-network is to estimate the state-action value function and assist in decision-making, which we categorize into two types. \textit{Type I}, illustrated by the top-left of figure \ref{fig:qnetwork}, employs a classification neural network as the Q-network \cite{mnih2013playing}. It takes the identical observations or extracted features as input and generates a $|A|$-dimensional vector, where $|A|$ is the size of the action space. Each element represents an estimate of the value function of each action given state. 
Consequently, they face challenges when the action space expands, as in routing scenarios with router addition. This is because they necessitate expanding the output category through substantial retraining, thus causing difficulties in adapting effectively. A similar issue arises in the policy network, affecting the model's adaptability to dynamic topology, as observed in the PPO algorithm \cite{schulman2017proximal}. \textit{Type II}, captured by the top-right of figure \ref{fig:qnetwork} and often used in critic-network such as MADDPG \cite{lowe2017multi}, typically takes observations combined with specific action values as input, outputting a single Q value. Discrete actions are often encoded using enumeration. Likewise, introducing a new action type into the action space causes these networks to encounter unseen enumerative action values as input, where input distribution shifts and estimating the value function requires retraining. Therefore, traditional Q-networks are challenging to apply in scenarios where the routing topology changes.

For our proposed Q-network score layer, agent $i$ evaluates the value function separately based on its own feature and the features from neighbors, as portrayed by the bottom of figure \ref{fig:qnetwork}. These features are extracted through the preceding observation encoding and emergent communication layers, which capture the knowledge about its local subtopology and align the feature space. Unlike \textit{Type I}, which calculates concurrently for all actions, it performs a separate forward operation of the neural network for each neighbor to obtain Q values. This ensures dimensional input-output alignment and enhances the action space's scalability. Moreover, in contrast to \textit{Type II}, our Q-network consistently deals with known input distributions due to feature alignment, guaranteeing the stability and reliability of the model calculation in the dynamic topology. Therefore, our q-network could adaptively evaluate the long-term reward for transmitting to the new neighboring router without retraining.

Specifically, we concatenate the features extracted by router $i$ and its adjacent router $j$ and then pass them through the non-linear neural networks $F_3$. 
The value function of router $i$ transmitting a packet to router $j$ is calculated as:
\begin{equation}
    Q_{i,j}=F_3(\text{concat}(f_i, f_j))
    \label{eq:qq}
\end{equation}
After obtaining the value functions of each neighbor, the router could select one with the highest value for packet forwarding.

\subsection{Model Training}
The training procedure of DRAMA follows the DQN\cite{mnih2013playing}, and during the training phase, a replay buffer is maintained. Specifically, at each timestep, a tuple $<\mathcal{O}, \mathcal{A}, \mathcal{O}^{\prime}, \mathcal{R}, \mathcal{D}>$ is stored, where $\mathcal{O}, \mathcal{A}, \mathcal{O}^{\prime}, \mathcal{R}$ denote the set of the current joint observations, joint action, next joint observations and joint rewards for all $N$ routers, respectively.
Moreover, $\mathcal{D}=\{d_1, d_2, ..., d_N\}$ indicates whether packets need to continue forwarding. If router $i$ forwards a packet that loses or reaches its destination at the next hop, then $d_i=1$; otherwise, $d_i=0$. Every $u$ timesteps, the model selects a batch with size $\mathrm{B}$ for training. Then, the temporal difference (TD) loss between the predicted Q-values and the target values is defined as:
\begin{equation}
\mathcal{L_{TD}}(\theta) = \frac{1}{\mathrm{B}} \sum_{\mathrm{B}} \frac{1}{\mathrm{N}} \sum_{i=1}^{\mathrm{N}} \left(Q_{i,j}(\mathcal{O}; \theta) - y_i\right)^2
\label{eq:loss1}
\end{equation}

Here, router $j$ represents the action $a_i$ decided by router $i$, indicating that router $i$ transmits the packet to the next-hop router $j$. Note that the Q-value $Q_{i, j}$ is only related to the C-hop neighbors of nodes $i$ and $j$, not all observations O. The target value is computed as $y_i=r_i + (1-d_i)\gamma \max _{a^{\prime}\in n_j} Q_{j, a^{\prime}}(\mathcal{O}^{\prime};\theta^{*})$, where $n_j$ represents the neighbor set for router $j$. The Q-network is parameterized by $\theta$, and $\theta^{*}$ denotes the parameters of the target network. We perform a soft update on the target network parameters, i.e., $\theta^{*}=\beta \theta+(1-\beta) \theta^{*}$, where $\beta$ is a coefficient in the range 0 to 1.

Moreover, to facilitate the convergence of the neural network in complex network topologies, we also employed estimated cost (EC) loss as an additional constraint:
\begin{equation}
\mathcal{L_{EC}}(\theta) = \frac{1}{\mathrm{B}} \sum_{\mathrm{B}} \frac{1}{\mathrm{N}} \sum_{i=1}^{\mathrm{N}} \left(Q_{i,j}(\mathcal{O}; \theta) - WSP(i,j,z, \mathcal{O})\right)^2
\label{eq:loss2}
\end{equation}
where $WSP(i,j,z, \mathcal{O})$ is the weighted shortest path following the weighted $Cost_{i,j}$ if router $i$ transmits the packet through the neighbor $j$ to the destination $z$, and the queue of each router in the network keeps as observed in $\mathcal{O}$. The weighted cost $Cost_{i,j}$ of the link from router $i$ to router $j$ is the sum of the length of the link ($l_l$) between $i$, $j$ and the length of queue ($l_q$) in router $j$. Finally, the training aims to minimize the loss:
\begin{equation}
\mathcal{L} = \mathcal{L_{TD}} + \mathcal{L_{EC}}
\label{eq:loss}
\end{equation}

\section{experiment}
\subsection{Experimental Setup}

\begin{figure}[t]
\centerline{\includegraphics[width=.78\linewidth]{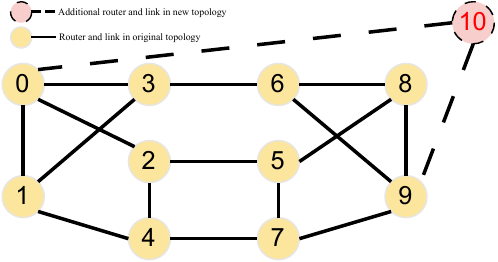}}
\caption{Network Topology Used In Simulator.}
\label{fig:topo}
\end{figure}

\textbf{Simulator Environment.} We implemented a Python simulation environment employing the topology depicted in figure \ref{fig:topo}. Each router is equipped with a buffer queue of a maximum length of 50. To simplify the simulation, each link has a constant bandwidth, and its length is set to cover the distance within one timestep unless otherwise specified. Each step is set to 1.0ms. At each timestep, we generate packets following a Poisson distribution with an average rate of $\lambda$. Each packet is randomly generated from routers \{0, 1, 2, 3, 4\} and sent to routers \{8, 9\}. For the reward, $r_{lost}$, $\tau_l$, and $\tau_q$ are set to 100, 1, and 1, respectively. In this setting, a higher delivery rate is prioritized and rewarded, leading to more advanced strategies that may sacrifice some latency for better delivery rates.

We evaluate the performance using the average delivery rate and latency of all packets in the network within a time interval. It is important to note that latency calculations consider a packet only upon arrival at its destination. For each experiment, 10 simulations with 512 timesteps are conducted, and the average value is presented as the result. Across all tables in this paper, the best results are highlighted in bold, and the second-best are highlighted with underlines.

\textbf{Baseline.} We selected six classic and state-of-the-art packet-level algorithms that are not confined to any specific network as baselines for comparison: the Shortest Path First algorithm (SPF), the Backpressure \cite{tassiulas1990stability}, the centralized RL algorithm PPO \cite{schulman2017proximal}, the distributed RL algorithm Q-routing \cite{boyan1993packet}, the fully distributed MARL algorithm MADDPG \cite{lowe2017multi}, and the distributed MARL algorithm with conventional communication DQRC \cite{you2020toward}.

\begin{itemize}
    \item SPF: In the SPF algorithm, packets are transmitted through the next-hop router with the shortest path length to the destination, which is calculated in advance.
    \item Backpressure (BP): The Backpressure algorithm is a dynamic, queue-based method that maximizes network throughput based on the differential backlog of packets at nodes, effectively balancing load and avoiding congestion.
    \item Q-routing: In the Q-routing algorithm, tabular Q-values are used to represent the estimated delivery time to destination nodes, optimizing network performance through experience-based Q-learning.
    \item PPO: The PPO algorithm is a state-of-the-art single-agent reinforcement learning algorithm. It assumes the presence of a central controller capable of obtaining observations from all routers and making decisions for each router.
    \item MADDPG: 
    In MADDPG, each intelligent agent has an actor network and a critic network, both consisting of three fully connected layers. Because the MADDPG algorithm does not involve communication between routers, each router makes distributed decisions based solely on its own observations during deployment.
    \item DQRC: In DQRC, each router leverages the LSTM as a decision network for forwarding packets individually, and they could share the raw queue length to perceive dynamic change.
\end{itemize}

\textbf{Implementation Details.} All reinforcement learning algorithms set GAMMA to 0.99 and use a soft update parameter, $\beta$, set to 0.01. For DRAMA, $F_1$ and $F_2$ are two-layer fully connected networks with ReLU, and sigmoid activation is leveraged for output to scale messages within $(0,1)$. $F_3$ is a two-layer fully connected network with ReLU. All hidden layer and message dimensions are set to 8, and batch normalization \cite{ioffe2017batch} is used to accelerate the learning process. The maximum communication rounds $C$ is 2 for the emergent communication layer. Moreover, the self-attention scaling factor, $\tau$, is set to 0.25, and a dropout of 0.3 is applied to attention coefficients. 

\subsection{The Performance Analysis of Static Network Topology}
We initially evaluate the performance of each algorithm on a static topology.

\textbf{Network Load.} We present the results of testing the algorithms under varying network loads in table \ref{tab:load}. Under low network load conditions ($\lambda$=1), all algorithms achieve a 100\% delivery rate, and except Backpressure, they all exhibit similar low latency. However, due to its inability to perceive router queuing congestion, the SPF algorithm and q-routing experience a significant decrease in both delivery rate and latency as the network load increases, falling behind other algorithms. It is important to note that most traditional algorithms cannot dynamically adjust routing rules based on network performance, as observed in SPF, making them less effective in addressing network congestion. Moreover, while the MADDPG demonstrates some performance advantages over SPF and Q-routing when $\lambda$$>$1 because of central training, it is evident that it has a 20\% loss rate when $\lambda$ reaches 4. Since MADDPG makes decisions based solely on local observations, it is prone to getting stuck in the local optima instead of reaching the global optima. Furthermore, because of the global sharing of information, PPO could achieve almost 100\% delivery rate when $\lambda=$3 and the lowest latency except for DRAMA. However, it could only reach an 89.6\% arrival rate when $\lambda=$4 due to the complex optimization task. Notable, although Backpressure experiences bad performance for low $\lambda$, it performs better when the network becomes congested by using congestion gradients.

DRAMA and DRQC algorithms achieve the best delivery rate as network congestion increases due to routers' communication. However, DRAMA could guarantee 100\% packet delivery when $\lambda$=4 and only requires 36.92\% of the average delivery time of DRQC. This is because the learnable emergent communication, rather than predefined fixed message content, could be self-organized to overcome the extreme network conditions. Furthermore, DRAMA can ensure the highest delivery rate and lowest latency across all network loads. These observations demonstrate that DRAMA with emergent communication outperforms the six baselines in various network loads.

\begin{table}[]
\centering
\caption{Results under different network load parameters ($\lambda$). DRAMA- denotes the DRAMA trained only by TD loss.}
\label{tab:load}
\resizebox{\columnwidth}{!}{%
\begin{tabular}{|c|c|c|c|c|c|}
\hline
                               & Algorithm    & $\lambda$=1                  & $\lambda$=2                & $\lambda$=3                & $\lambda$=4                 \\ \hline
\multirow{8}{*}{Delivery Rate} & SPF          & 1.0000               & 1.0000             & 0.9395             & 0.8236              \\ \cline{2-6} 
                               & BP & 1.0000               & 1.0000             & \textbf{1.0000}    & 0.9735              \\ \cline{2-6} 
                               & Q-routing    & 1.0000               & 1.0000             & 0.9499             & 0.8647              \\ \cline{2-6} 
                               & MADDPG       & 1.0000               & 1.0000             & 0.9697             & 0.8010              \\ \cline{2-6} 
                               & PPO          & 1.0000               & 1.0000             & {\ul 0.9992}       & 0.8963              \\ \cline{2-6} 
                               & DRQC         & 1.0000               & 1.0000             & \textbf{1.0000}    & 0.9431              \\ \cline{2-6} 
                               & DRAMA-       & 1.0000               & 1.0000             & \textbf{1.0000}    & {\ul 0.9878}        \\ \cline{2-6} 
                               & DRAMA        & 1.0000               & 1.0000             & \textbf{1.0000}    & \textbf{1.0000}     \\ \hline
\multirow{8}{*}{Latency(ms)}       & SPF          & {\ul 3.33{\tiny±0.20}}      & 15.42{\tiny±5.10}         & 42.40{\tiny±4.23}         & 50.61{\tiny±3.89}          \\ \cline{2-6} 
                               & BP & 6.45{\tiny±0.21}             & 7.78{\tiny±0.30}          & 17.59{\tiny±3.54}         & 54.94{\tiny±3.29}          \\ \cline{2-6} 
                               & Q-routing    & 3.37{\tiny±0.22}            & 12.86{\tiny±5.30}         & 34.77{\tiny±3.23}         & 55.03{\tiny±2.23}          \\ \cline{2-6} 
                               & MADDPG       & 3.56{\tiny±0.23}            & 5.08{\tiny±1.21}          & 26.19{\tiny±2.56}         & 36.14{\tiny±10.45}         \\ \cline{2-6} 
                               & PPO          & 4.18{\tiny±0.07}            & 5.06{\tiny±0.89}          & 16.02{\tiny±2.34}         & 41.02{\tiny±6.53}        \\ \cline{2-6} 
                               & DRQC         & 3.55{\tiny±0.06}            & 4.43{\tiny±0.69}          & 20.31{\tiny±2.83}         & 49.15{\tiny±13.05}         \\ \cline{2-6} 
                               & DRAMA-       & 3.53{\tiny±0.09}            & {\ul 4.35{\tiny±0.12}}    & {\ul 5.79{\tiny±1.41}}    & {\ul 19.52{\tiny±2.27}}    \\ \cline{2-6} 
                               & DRAMA        & \textbf{2.81{\tiny±0.04}} & \textbf{4.04{\tiny±0.62}} & \textbf{5.34{\tiny±1.01}} & \textbf{18.15{\tiny±3.72}} \\ \hline
\end{tabular}
}
\end{table}

\begin{figure}[t]
\centerline{\includegraphics[width=.6\linewidth]{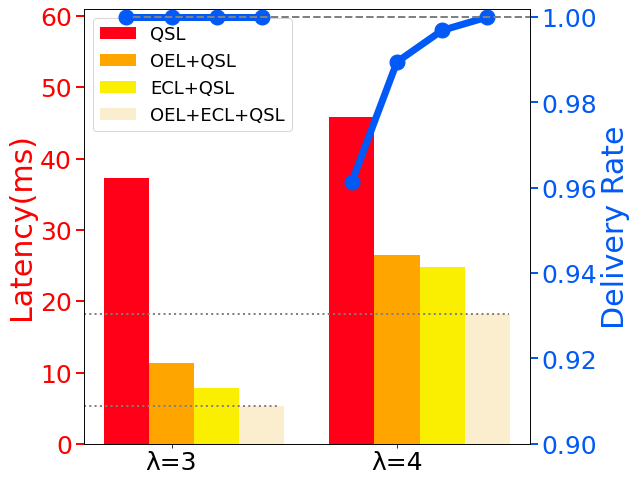}}
\caption{Ablation Experiments}
\label{fig:ablation}
\end{figure}

\textbf{Ablation.} In figure \ref{fig:ablation}, we perform various ablations. Within DRAMA, the Q-network score layer is indispensable. A model with only the Q-network score layer (QSL) can be considered as engaging in a type of communication that merely shares observations. Meanwhile, a model comprising both the observation encoder layer and the Q-network score layer (OEL+QSL) can be perceived as a fundamental instance of emergent communication among routers, where the OEL plays the role of a learnable message encoder. Finally, ECL+QSL is the model without the observation encoding layer, thus observation is leveraged directly.

The experiment results show that the QSL performs weakly in both delivery rate and latency metrics compared to the OEL+QSL and ECL+QSL. This suggests that effective collaboration for packet forwarding necessitates not only sharing observations but also a deeper level of communication among routers, which is enhanced by emergent communication. Therefore, emergent communication is beneficial to the efficiency of packet routing by intelligently exchanging task-specific information.

Additionally, our complete DRAMA model (OEL + ECL + QSL), with messages exchanged by the encoder and attention mechanism, captures knowledge from more distant routers, expanding the scope and depth of negotiation and collaboration, thereby achieving remarkable performance.

\begin{figure}[t]
\centerline{\includegraphics[width=.6\linewidth]{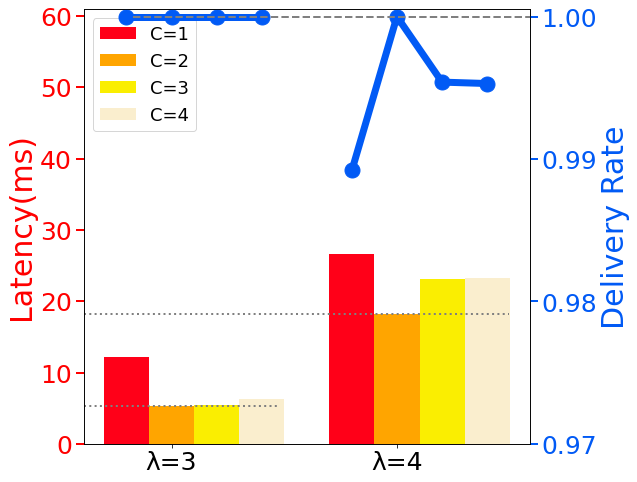}}
\caption{Results of Communication Round Experiments}
\label{fig:round}
\end{figure}

\textbf{Communication Round.} In Figure \ref{fig:round}, we investigated the impact of the communication rounds in the emergent communication layer $C$. The model with $C$=2 gains a higher delivery rate and lower latency than that with $C$=3 and 4. This might be attributed to the smoothing effect of the feature or overfitting due to the complexity of the model. Moreover, compared to models with $C$=1, the model with $C$=2 demonstrates a relative latency decrease of approximately 56\% at $\lambda$=3 and an absolute increase in delivery rate of approximately 1.1\% at $\lambda$=4. This suggests that an appropriately large number of communication rounds could promote collaborative packet forwarding.

\textbf{Communication Overhead.} In table \ref{tab:quantization}, we report the overhead of the emergent communication in DRAMA and demonstrate that only two simple strategies are needed to reduce overhead. The first strategy is to quantify the shared message from 4 bytes (float32) to 1 bit, which is denoted as "-Q" in the table. The second strategy is to share the message per 10 steps, which is denoted as "-H10". Since the message dimension is 8, the message size generated by each timestep and router is 256 bits, and so on for others. It is shown that with these two strategies, DRAMA-Q-H10 only loses about 0.69\% delivery rate and 3ms latency while using only 0.3\% of the communication volume compared with original DRAMA, when $\lambda$=4.

\begin{table}[]
\centering
\caption{Results of Communication Overhead}
\label{tab:quantization}
\resizebox{\columnwidth}{!}{%
\begin{tabular}{|c|c|c|c|c|}
\hline
\multicolumn{1}{|l|}{}       &             & Delivery Rate & Latency(ms)    & Overhead(bits) \\ \hline
\multirow{4}{*}{$\lambda$=3} & DRAMA       & 1.0000        & 5.34{\tiny±1.01}  & 256     \\ \cline{2-5} 
                             & DRAMA-Q     & 1.0000        & 7.67{\tiny±1.37}  & 8       \\ \cline{2-5} 
                             & DRAMA-H10   & 1.0000        & 6.88{\tiny±1.46}  & 25.6    \\ \cline{2-5} 
                             & DRAMA-Q-H10 & 1.0000        & 7.72{\tiny±1.86}  & 0.8     \\ \hline
\multirow{4}{*}{$\lambda$=4} & DRAMA       & 1.0000        & 18.15{\tiny±3.72} & 256     \\ \cline{2-5} 
                             & DRAMA-Q     & 0.9988        & 21.33{\tiny±1.86} & 8       \\ \cline{2-5} 
                             & DRAMA-H10   & 0.9997        & 21.34{\tiny±3.08} & 25.6    \\ \cline{2-5} 
                             & DRAMA-Q-H10 & 0.9931        & 22.17{\tiny±2.88} & 0.8     \\ \hline
\end{tabular}
}
\end{table}

\subsection{The Performance Analysis of Dynamic Network Topology}
We evaluated the adaptability of DRAMA to dynamic network topology.

\begin{figure}[t]
    \centering 
    \subfigure[Result of Link Failure] { 
    \label{fig:linke_failure} 
     \includegraphics[width=0.42\linewidth]{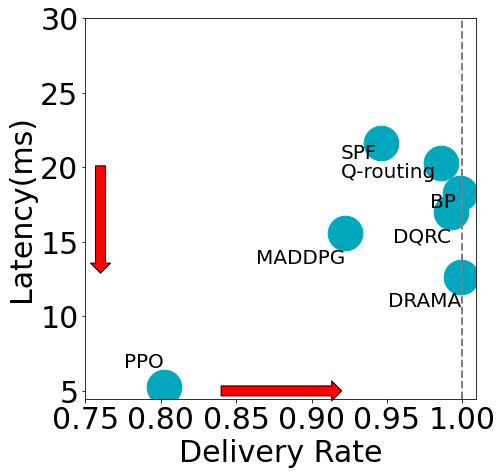}}
     % \hfill
     \subfigure[Result of Node Failure] { 
     \label{fig:node_failure}
    \includegraphics[width=0.42\linewidth]{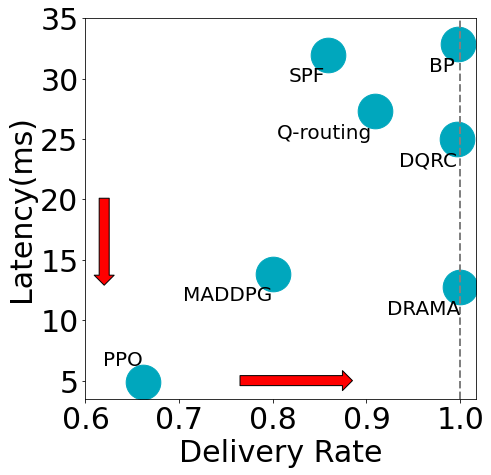}}
    \caption{The Result of Failure}
    \label{fig:failure}
    \vspace{-4mm}
\end{figure}

\textbf{Link Failure and Node Failure.} The failure of network links is a common reason that causes changes to the network topology. Figure \ref{fig:linke_failure} shows the performance of each routing algorithm encountering link failures. Due to constrained link resources, $\lambda$ is set to 2. During testing, a random link is chosen to invalidate each simulation episode. In this experiment, we performed 50 random simulations for evaluation. 

For the traditional algorithm, SPF could not achieve an optimal arrival rate and latency, while BP could reach a high arrival rate but with a high latency. These algorithms make it hard to develop different strategies based on topological changes. It is exhibited that although the latency of PPO and MADDPG are low, especially only 5ms for PPO, their delivery rate cannot be guaranteed, only about 80\% and 90\%, respectively. Moreover, Q-routing takes more timesteps to ensure a high delivery rate. This indicates a significant deficiency in existing RL/MARL-based algorithms without communication for dynamic network topology. They are sensitive to topology changes, possibly due to these algorithms not capturing the structural features of the network.
In contrast, DRAMA and DQRC could achieve almost 100\% arrival rate with communication. However, DRAMA could gain lower latency, achieving a Pareto optimal solution and striking a balance between latency and delivery rate. This is because, through emergent communication among routers, routers could self-organize their communication and perceive the network state and topology changes in real time when the link fails, thereby ensuring adaptive decision-making. Similar observations could be found in figure \ref{fig:node_failure}, where we randomly selected one router from the set \{5, 6, 7\} to undergo failure and set $\lambda$ to 2. Therefore, DRAMA could adapt well to dynamic changes in topology caused by router and link failure.

\begin{table}[]
\centering
\caption{Result of Node Addition}
\label{tab:addnode}
\resizebox{\columnwidth}{!}{%
\begin{tabular}{|c|cc|cc|}
\hline
\multicolumn{1}{|l|}{} & \multicolumn{2}{c|}{$\lambda$=3}                   & \multicolumn{2}{c|}{$\lambda$=4}                   \\ \hline
\multicolumn{1}{|l|}{} & \multicolumn{1}{c|}{Delivery Rate} & Latency(ms)      & \multicolumn{1}{c|}{Delivery Rate} & Latency(ms)      \\ \hline
SPF                    & \multicolumn{1}{c|}{{\ul 0.9891}}        & 16.83{\tiny±1.12}  & \multicolumn{1}{c|}{0.9174}        & 41.63{\tiny±1.88}  \\ \hline
BP           & \multicolumn{1}{c|}{\textbf{1.0000}}        & 7.22{\tiny±0.41}  & \multicolumn{1}{c|}{\textbf{1.0000}}        & 16.67{\tiny±1.51} \\ \hline
Q-routing              & \multicolumn{1}{c|}{\textbf{1.0000}}        & 14.79{\tiny±3.16}  & \multicolumn{1}{c|}{0.9255}        & 28.49{\tiny±2.92}  \\ \hline
MADDPG                 & \multicolumn{1}{c|}{\textbf{1.0000}}        & {16.39{\tiny±1.37}}  & \multicolumn{1}{c|}{\textbf{1.0000}}        & {38.09{\tiny±1.72}}  \\ \hline
PPO                    & \multicolumn{1}{c|}{\textbf{1.0000}}        & 6.41{\tiny±0.98} & \multicolumn{1}{c|}{{\ul 0.9961}}        & {19.75{\tiny±2.13}} \\ \hline
DQRC                   & \multicolumn{1}{c|}{\textbf{1.0000}}        & {\ul 6.21{\tiny±1.53}} & \multicolumn{1}{c|}{\textbf{1.0000}}        & {\ul 16.01{\tiny±2.74}} \\ \hline
DRAMA                  & \multicolumn{1}{c|}{\textbf{1.0000}}        & \textbf{4.48{\tiny±0.61}}    & \multicolumn{1}{c|}{\textbf{1.0000}}        & \textbf{14.47{\tiny±1.43}}  \\ \hline
\end{tabular}
}
\end{table}

\begin{table}[]
\centering
\caption{Result in Static ATT}
\label{tab:att}
\begin{tabular}{|c|c|c|}
\hline
\multicolumn{1}{|c|}{} & Delivery Rate & Latency(ms)     \\ \hline
SPF                    & {\ul 0.9912}        & {\ul 19.24{\tiny±17.38}} \\ \hline
BP           & 0.9552        & 99.02{\tiny±29.11} \\ \hline
Q-routing              & 0.6218        & 62.61{\tiny±10.94} \\ \hline
MADDPG                 & 0.3127        & 26.68{\tiny±9.04}  \\ \hline
PPO                    & 0.5100        & 227.87{\tiny±9.48} \\ \hline
DQRC                   & 0.4831        & 132.31{\tiny±9.33} \\ \hline
DRAMA                  & \textbf{1.0000}        & \textbf{5.94{\tiny±1.46}}  \\ \hline
\end{tabular}
\end{table}

\begin{table}[]
\centering
\caption{Result of Link Failure in ATT}
\label{tab:failure12}
\resizebox{\columnwidth}{!}{%
\begin{tabular}{|c|cc|cc|}
\hline
\multicolumn{1}{|l|}{} & \multicolumn{2}{c|}{failure\#1}                   & \multicolumn{2}{c|}{failure\#2}                   \\ \hline
\multicolumn{1}{|l|}{} & \multicolumn{1}{c|}{Delivery Rate} & Latency(ms)      & \multicolumn{1}{c|}{Delivery Rate} & Latency(ms)      \\ \hline
SPF                    & \multicolumn{1}{c|}{{\ul 0.9901}}        & 28.42{\tiny±27.79}  & \multicolumn{1}{c|}{{\ul 0.9888}}        & 29.18{\tiny±28.24}  \\ \hline
BP           & \multicolumn{1}{c|}{0.9592}        & 93.28{\tiny±30.59}  & \multicolumn{1}{c|}{0.9413}        & 103.22{\tiny±28.77} \\ \hline
Q-routing              & \multicolumn{1}{c|}{0.6134}        & 59.33{\tiny±11.54}  & \multicolumn{1}{c|}{0.5917}        & 59.06{\tiny±13.24}  \\ \hline
MADDPG                 & \multicolumn{1}{c|}{0.2418}        & {\ul 15.68{\tiny±10.75}}  & \multicolumn{1}{c|}{0.2321}        & {\ul 27.23{\tiny±12.68}}  \\ \hline
PPO                    & \multicolumn{1}{c|}{0.2025}        & 222.28{\tiny±10.41} & \multicolumn{1}{c|}{0.1737}        & 197.59{\tiny±14.12} \\ \hline
DQRC                   & \multicolumn{1}{c|}{0.2953}        & 110.84{\tiny±21.13} & \multicolumn{1}{c|}{0.2838}        & 117.48{\tiny±31.06} \\ \hline
DRAMA                  & \multicolumn{1}{c|}{\textbf{0.9994}}        & \textbf{7.05{\tiny±3.04}}    & \multicolumn{1}{c|}{\textbf{0.9981}}        & \textbf{10.03{\tiny±10.41}}  \\ \hline
\end{tabular}
}
\end{table}

\begin{table}[]
\centering
\caption{Result of Router Extension in ATT}
\label{tab:attadd}
\begin{tabular}{|c|c|c|}
\hline
      & \multicolumn{1}{c|}{Delivery Rate} & \multicolumn{1}{c|}{Latency(ms)} \\ \hline
SPF   & {\ul 0.9995}                             & {\ul 9.15{\tiny±12.91}}                  \\ \hline
BP    & 0.9834                             & 65.07{\tiny±27.53}                  \\ \hline
DRAMA & \textbf{1.0000}                                  & \textbf{4.57{\tiny±2.18} }                   \\ \hline
\end{tabular}
\end{table}

\textbf{Add Routers.} The performance of DRAMA in scenarios involving the addition of the router is evaluated in table \ref{tab:addnode}, where a new router connecting router 0 and router 9 is introduced, thus forming a new network topology as shown in figure \ref{fig:topo}. As analyzed in the previous section, MADDPG, DQRC, PPO, and Q-routing have inherent limitations in their Q-network/table structures, which render them incapable of supporting the addition of nodes. Consequently, these algorithms undergo training directly on the new topology. By ensuring the alignment of latent features and input-output between the old and new routers through the proposed ECL and QSL, DRAMA could adapt to changes in the number of neighbors without retraining. 

Compared with table \ref{tab:load}, the performance of each algorithm increases. Because of inflexibility, SPF and Q-routing still could not transmit reliably, having only 91.74\% and 92.55\% delivery rates when $\lambda$=4. Moreover, BP, DQRC, and PPO learn to employ router 10 as a shortcut for fast transmission, thus achieving good latency with about an average of 16ms and 19ms timesteps. However, the existing RL/MARL algorithms in the baseline necessitate training in the new topology again to leverage the added router. 

DRAMA experiences superior utilization of the added routers for packet routing directly, reaching an average of 4.48ms when $\lambda$=3 and 14.47ms when $\lambda$=4. This means DRAMA could perceive the real-time conditions of the current network rather than fitting precisely to a particular topology, and the high generalization capability of DRAMA is underscored. Therefore, DRAMA could scale to new routers, achieving high adaptability and efficiency without additional training.

\subsection{Real-world Topology}

\begin{figure}[t]
\centerline{\includegraphics[width=.8\linewidth]{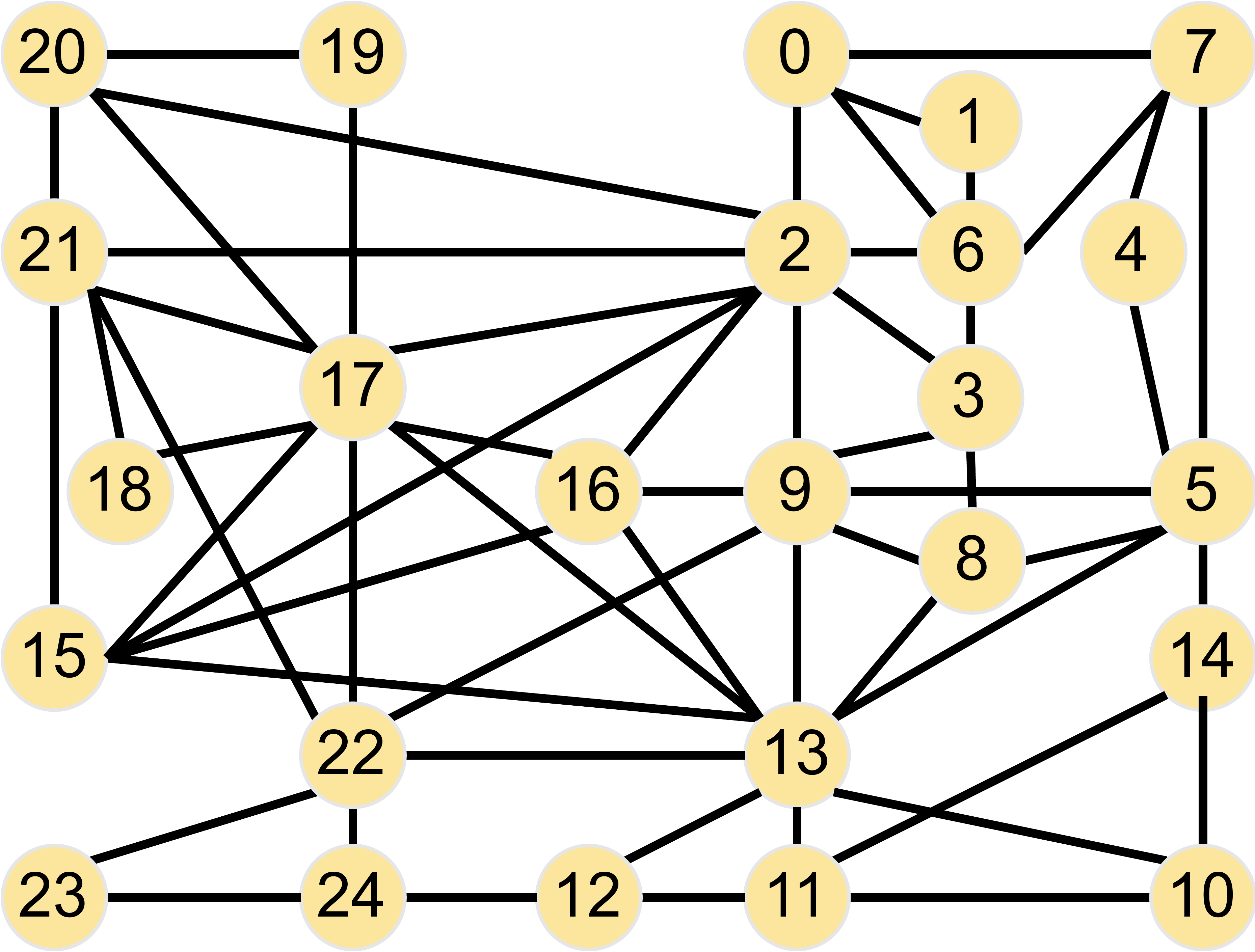}}
\caption{Topology of ATT}
\label{fig:att}
\end{figure}

We evaluate the algorithm in the real-world topology ATT \cite{kumar2018semi} and real-world flow demand, where the topology is as figure \ref{fig:att} and the performance is illustrated in table \ref{tab:att}. Because of the large number of routers and busy demands, the RL/MARL algorithms in the baseline are hard to train well. Consequently, they all perform worse than the traditional algorithm SPF and Backpressure. However, through emergent communication for cooperation and additional loss for promoting convergence, DRAMA still could transmit the packet at a 100\% arrival rate and only need about 1/3 of the latency of SPF. 

Furthermore, We randomly disable one (failure\#1) and two (failure\#2) links to demonstrate the adaptability to dynamic topology, shown in table \ref{tab:failure12}. PPO, DQRC, and MADDPG are sensitive to link failure, while others can adjust their transmission strategies and relieve packet loss. However, 99.8\% delivery rate and 10 timesteps latency are obtained by DRAMA, revealing the adaptability assisted by the emergence of communication.

For the experience of router extension, a router is inserted into two random routers in each simulation, and 50 simulations are conducted, shown in table \ref{tab:attadd}. Since the RL/MARL algorithms in baseline need to be retrained for each new topology with the additional router, we only evaluate our DRAMA and two traditional algorithms. BP still costs an average of about 65 timesteps to transfer the packets, while SPF only needs an average of about 9 timesteps. However, DRAMA only requires half the time of SPF. Therefore, DRAMA could outperform all baselines in topology ATT and demonstrate adaptability to dynamic topology.

\section{Conclusion and Future Work}
We proposed DRAMA as a novel algorithm employing MARL with emergent communication for packet routing in dynamic networks. Numerical results have demonstrated that DRAMA with emergent communication outperforms three types of baseline algorithms under various network statuses in toy and real-world topology. The ablation study and the communication round evaluation have validated the potential effect of emergent communication in packet routing. It is also shown that the novel Q-network and graph-based emergent communication could benefit DRAMA in enhancing collaboration among routers and adapting to dynamic router/link failure and router addition, achieving reliable and efficient packet transmission.

We point out three directions for future work. First, it is worth investigating the model's scalability by using a larger and richer network topology. Second, exploring the trade-off between communication overhead and task efficiency is a promising direction. Last, it is crucial to investigate the impact of communication delay on packet routing problems in real-world networks.

\section*{Acknowledgment}

This research is supported in part by the Huawei Technologies Co., Ltd., in part by the National Natural Science Foundation of China (Grant No. 92067109), in part by Shenzhen Science and Technology Program (Grant No. ZDSYS20210623092007023), and in part by the Guangdong Basic and Applied Basic Research Foundation (Grant No. 2021A1515110685)

\bibliographystyle{IEEEtran}
\bibliography{IEEEabrv,references}

\end{document}